\documentclass[A4,10pt]{iopart}
\usepackage{amsfonts}
\usepackage{geometry}
\usepackage{graphicx}

\begin{document}

\title{Understanding critical behavior in the framework of the extended equilibrium fluctuation theorem}
\author{L. Velazquez and S. Curilef}

\begin{abstract}
Recently, we have derived a fluctuation theorem for systems in thermodynamic equilibrium compatible with anomalous
response functions, e.g. the existence of states with \textit{negative heat capacities} $C<0$. In this work, we show
that the present approach of the fluctuation theory introduces new insights in the understanding of \textit{critical phenomena}.
Specifically, the new theorem predicts that the environmental influence can radically affect critical behavior of systems, e.g. to
provoke a suppression of the divergence of correlation length $\xi$ and some of its associated phenomena as spontaneous symmetry
breaking. Our analysis reveals that while response functions and state equations are \emph{intrinsic properties} 
for a given system, critical behaviors are always \emph{relative phenomena}, that is, their existence crucially depend on the underlying environmental influence. \newline
\end{abstract}

\address{Departamento de F\'\i sica, Universidad Cat\'olica del Norte, Av. Angamos 0610, Antofagasta,
Chile.}

\tableofcontents

\newpage
\section{Introduction}
\textit{Critical phenomena} is the collective name associated with
the physics of \textit{critical points} \cite{Stanley}, which is
originated as a consequence of the divergence of the correlation
length $\xi $. The understanding of critical phenomena has motivated
the development of new mathematical tools for the study of phase
transitions; such as the \textit{renormalization group theory},
which has a significant impact in condensed matter and high energy
physics \cite{Gold}. In this work, we present simple theoretical arguments and
simulations that reveal new aspects in the understanding of critical
phenomena. We show that \textit{the environmental influence can
radically affect the critical behavior} of systems, as example, to
provoke a suppression of the divergence of correlation length $\xi$
and some of its associated phenomena as spontaneous symmetry
breaking.

Our arguments are based on a framework of equilibrium fluctuation theory recently proposed \cite{vel.unc,vel.unc11,vel.unc12,vel.unc2,vel.unc3}. The main contributions of this approach are the derivation of a set of fluctuation theorems compatible with \emph{anomalous response functions} \cite{vel.unc,vel.unc11,vel.unc12,vel.unc2} as well as \emph{uncertainty relations involving thermodynamic quantities} \cite{vel.unc3}. For the sake of self-consistence of this paper, we firstly present a brief review of the so-called \emph{fundamental equilibrium fluctuation theorem} and some of its most immediate consequences \cite{vel.unc2}. After, this theorem will be considered to analyze the role of the environmental influence on critical phenomena. The results of this analysis will be illustrated studying the critical behavior 2D Ising model. Finally, we present some concluding remarks.

\section{Extended equilibrium fluctuation theorem}
The fundamental equilibrium fluctuation theorem (EFT):
\begin{equation}
\mathcal{R} =C+\mathcal{R} D  \label{efdt}
\end{equation}
describes the relation between the system response functions and its fluctuating behavior \cite{vel.unc2}. The quantity
$\mathcal{R} $ represents the \textit{response matrix}:
\begin{equation}
\mathcal{R} =-\left(
\begin{array}{cc}
\partial _{\beta }\mathcal{H} & \partial _{\beta }\left( \beta X\right) \\
\partial _{Y}\mathcal{H} & \beta \partial _{Y}X%
\end{array}%
\right) ,  \label{response}
\end{equation}
which characterizes the system response to the environmental
influence; whereas $Y=\left( p,-\vec{E},-\vec{H},\ldots \right) $
denotes a generalized force (pressure, electric field, magnetic
field, etc.) and $ X=\left( V,\vec{P},\vec{M},\ldots \right) $ the
corresponding generalized displacement (volume, polarization,
magnetization, etc.). Besides, $\mathcal{H}=U+YX$ and $U$ are the
system Enthalpy and the internal energy, respectively; while $\beta
=1/T$\footnote{Boltzmann constant $k_{B}$ has been set as the
unity.} denotes the inverse temperature and $\partial _{x}A$, the
partial derivative $\partial A/\partial x$. The quantity $C$ is the
\textit{ self-correlation matrix} that characterizes the system
fluctuating behavior:
\begin{equation}
C=\left(
\begin{array}{cc}
\left\langle \delta Q^{2}\right\rangle & \beta \left\langle \delta
Q\delta
X\right\rangle \\
\beta \left\langle \delta X\delta Q\right\rangle & \beta
^{2}\left\langle
\delta X^{2}\right\rangle%
\end{array}%
\right) ,  \label{self.correlation}
\end{equation}%
while the \textit{correlation matrix} $D$:
\begin{equation}
D=\left(
\begin{array}{cc}
\left\langle \delta \beta \delta Q\right\rangle & \beta \left\langle
\delta
\beta \delta X\right\rangle \\
\left\langle \delta Y\delta Q\right\rangle & \beta \left\langle
\delta
Y\delta X\right\rangle%
\end{array}%
\right)  \label{inter.correlation}
\end{equation}%
describes the existence of \textit{environmental feedback effects}
among the system macroscopic observables $\left( U,X\right) $ and
the environmental control variables $\left( \beta ,Y\right) $ due to
the underlying thermodynamic interaction. As expected, the amount of
heat exchanged between the system and the environment at the
equilibrium $\delta Q=\delta U+Y\delta X=T\delta S$ obeys the
condition $\left\langle \delta Q\right\rangle =0$.

EFT (\ref{efdt}) represents a suitable extension of the usual
fluctuation theorem \cite{Reichl}:
\begin{equation}
\mathcal{R} =C  \label{oefdt}
\end{equation}%
derived from the Boltzmann-Gibbs (BG) distributions:
\begin{equation}
dp_{BG}\left( \left. U,X\right\vert \beta ,Y\right) =\frac{\exp
\left[ -\beta \left( U+YX\right) \right] }{Z\left( \beta ,Y\right)
}\Omega(U,X)dUdX. \label{BG}
\end{equation}
Here, $\Omega(U,X)$ is the system density of states, while $Z\left( \beta ,Y\right)$ is the partition function.
Eq.(\ref{oefdt}), as example, contains the fluctuation relations:
\begin{equation}\label{usual}
C_{p}=\beta ^{2}\left\langle \delta Q^{2}\right\rangle
,~VK_{T}=\beta \left\langle \delta V^{2}\right\rangle ,~\chi
_{T}=\beta \left\langle \delta M^{2}\right\rangle,
\end{equation}%
where $C_{p}=T\left( \partial S/\partial T\right) _{p}$ is the
isobaric heat capacity, $K_{T}=-V^{-1}\left(
\partial V/\partial p\right) _{T}$ is the isothermal compressibility
and $\chi _{T}=\left( \partial M/\partial H\right) _{T}$ is the
isothermal magnetic susceptibility. Despite the wide application in
statistical mechanics, BG distributions (\ref{BG}) have a restricted
applicability due to the environmental inverse temperature $\beta $
and its generalized force $Y$ are regarded as \textit{constant
parameters}. Such that restriction demands the consideration of a
thermal contact with a bath with infinite heat capacity or a
reservoir with an infinite number of particles, as other
\textit{idealizations} assumed in conventional applications (usually
attributed to the natural environment). In a general equilibrium
situation, the internal state of the system acting as ``environment"
is also perturbed by the underlying thermodynamic interaction. This
is the origin of environmental feedback effects described by the
correlation matrix $D$ in the extended EFT (\ref{efdt}).

Conventional EFT (\ref{oefdt}) is only compatible with states with a positive definite response matrix $\mathcal{R}$. Considering its particular fluctuation relations (\ref{usual}), this last requirement implies the positive character of the heat capacity $C_{p}$, the isothermal compressibility $K_{T}$ and the magnetic susceptibility $\chi_{T}$. Remarkably, thermodynamics also supports the existence of states with \emph{negative heat capacities} and other  anomalous response functions, that is, a non-positive definite response matrix $\mathcal{R}$. Mathematically, the existence of anomalous response is associated with the presence of states where the entropy $S=\log W$ is a \emph{non-concave function}. Such a relationship comes from the possibility of expressing response matrix $\mathcal{R}$ as follows:
\begin{equation}
\mathcal{R}=-T\mathrm{H}^{-1}T^{T},
\end{equation}
where $\mathrm{H}$ is the entropy Hessian:
\begin{equation}
\mathrm{H}=\left(%
\begin{array}{cc}
  \partial^{2}_{U}S & \partial_{U}\partial_{X}S \\
  \partial_{X}\partial_{U}S & \partial^{2}_{X}S \\
\end{array}%
\right),
\end{equation}
and the transformation matrix $T$:
\begin{equation}
T=\left(%
\begin{array}{cc}
  1 & Y \\
  0 & \beta \\
\end{array}%
\right).
\end{equation}
Moreover, the thermodynamic quantities $\beta$ and $Y$ are also obtained from the entropy $S$ via the expressions:
\begin{equation}\label{var.deriv}
\beta=(\partial S/\partial U)_{X},\, \beta Y=(\partial S/\partial X)_{U}.
\end{equation}
The states with anomalous response conform the so-called regions of \emph{ensemble inequivalence} \cite{vel.unc,vel.unc11,vel.unc12,vel.unc2}. Physically, the anomalous response is a consequence of the incidence of \emph{non-extensive effects} \cite{vel.unc,vel.unc11,vel.unc12,vel.unc2}, as example, \emph{the presence of long-range interactions} or \emph{the development of spatial non-homogeneities during the occurrence of discontinuous phase transitions} \cite{Gross,thir,Dauxois}. The presence of a long-range force as gravitation explains the existence of negative heat capacities in astrophysical systems \cite{Gross,thir,Dauxois}. The development of \emph{interphases} induces the existence of \emph{surface correlations}, which are the origin of the negative heat capacities $C<0$ observed during the phase coexistence phenomenon \cite{Gross}. Analogously, such surface correlations explain the anomalous states with \emph{negative susceptibility} $\chi_{T}<0$ observed in a magnetic system below the critical temperature, which owe their origin to the formation of \emph{magnetic domains} \cite{vel.unc2}.

Direct consequences of the conventional EFT (\ref{oefdt}) are
known \cite{Reichl}:
\begin{itemize}
    \item The system fluctuating behavior is determined by its
    own response to the external conditions.

    \item The stable states are those ones where the response
    matrix $\mathcal{R}$ is also positive definite because of the
    self-correlation matrix $C$ is always positive definite.

    \item The simultaneous divergence of some components
    of the response matrix and their corresponding correlation
    functions at the critical points, $\mathcal{R}^{ij}\rightarrow\infty\Rightarrow
    C^{ij}\rightarrow\infty$, which is related to the divergence of
    correlation length, $\xi\rightarrow\infty$.
\end{itemize}
The presence of the correlation matrix $D^{ij}$ in the extended EFT
(\ref{efdt}) introduces the following radical changes in the
previous implications:
\begin{itemize}
    \item The system fluctuating behavior depends both on its
    response functions and the nature of the environmental
    influence.
    \item The stable states are those ones where $\mathcal{R}(I-D)$
    is a positive definite matrix, with $I$ being the identity
    matrix. States with anomalous response functions can be
    thermodynamically stable under certain environmental conditions.
    \item The divergence of the correlations functions $C^{ij}$ and
    the correlation length $\xi$ can be suppressed by the
    environmental influence despite the divergence of some components
    of the response matrix $\mathcal{R}^{ij}$ at the critical point.
\end{itemize}

The first and the second implications, together, claim that both \emph{the system fluctuating behavior and its thermodynamic stability depend on the nature of the environmental influence}. A trivial example is found in the case of the isolated system, where $\delta Q\equiv 0$. Clearly, there is no direct relation between the heat interchange and the heat capacity $C_{p}$ for this particular situation, which gives evidences about the restricted applicability of the fluctuation relation $C_{p}=\beta^{2}\left\langle\delta Q^{2}\right\rangle$. For a better understanding, let us consider a particular system where the unique relevant observable is the internal energy $U$. In this case, the EFT (\ref{efdt}) drops to the following energy-temperature fluctuation relation \cite{vel.unc}:
\begin{equation}\label{ETFR}
C=\beta^{2}\left\langle\delta
U^{2}\right\rangle+C\left\langle\delta\beta\delta U\right\rangle,
\end{equation}
which constitutes an extension of the fluctuation relation $C=\beta^{2}\left\langle\delta U^{2}\right\rangle$ obtained from the Gibbs' canonical ensemble (a special case of BG distributions):
\begin{equation}
dp_{G}(U|\beta)=\frac{\exp(-\beta U)}{Z(\beta)}\Omega(U)dU.
\end{equation}
Eq.(\ref{ETFR}) is compatible with the existence of states with negative heat capacities $C<0$ as long as the correlation function $\left\langle\delta\beta\delta U\right\rangle$ obeyed the stability condition:
\begin{equation}\label{stab}
\left\langle\delta\beta\delta U\right\rangle>1.
\end{equation}
The simplest scenario with non-vanishing correlated fluctuations, $\left\langle\delta\beta\delta U\right\rangle\neq 0$, is where the system is put in thermal contact with other short-range interacting system with a finite heat capacity $C^{e}$. In this case, the second system (environment) experiences a temperature fluctuation $\delta T=-\delta U/C^{e}$ whenever the system absorbs or releases the amount of energy $\delta U$. Substituting this last expression into Eq.(\ref{ETFR}), one obtains the self-correlation function of the internal energy $U$:
\begin{equation}
\frac{C^{e}C}{C^{e}+C}=\beta^{2}\left\langle\delta
U^{2}\right\rangle.
\end{equation}
Accordingly, the environmental heat capacity $C^{e}$ drives the system fluctuating behavior and imposes the standard microcanonical and canonical conditions in the asymptotic limits $C^{e}\rightarrow 0$ and $C^{e}\rightarrow\infty$, respectively. The stability condition (\ref{stab}) leads to the so-called Thirring's constraint $C^{e}<|C|$ necessary for the stability of negative heat capacities in the framework of short-range interacting systems \cite{thir}.

\section{Implications on critical phenomena}
\begin{figure}[t]
\begin{center}
\includegraphics[
width=4.0in ]{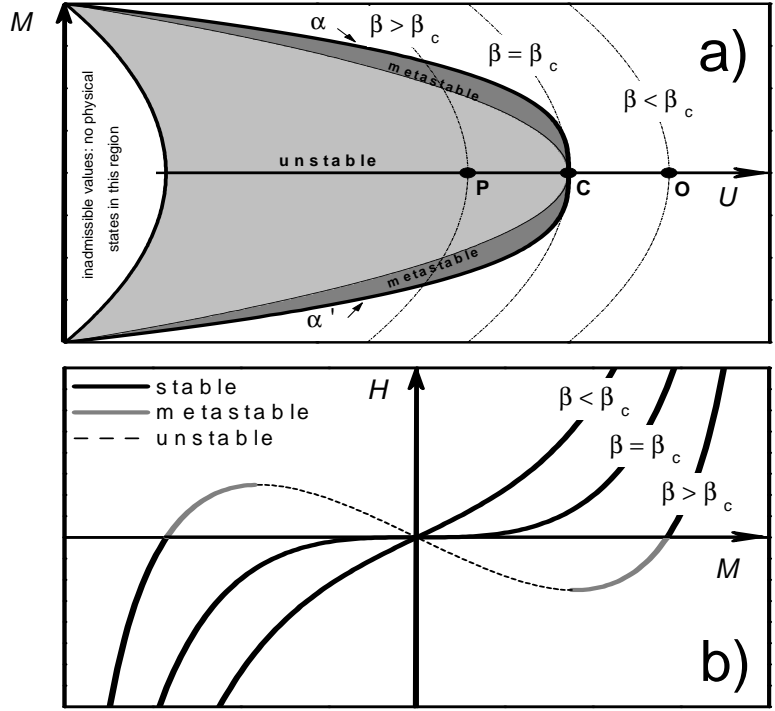}
\end{center}
\caption{Panel (a) Schematic representation of the regions of
thermodynamic stability of a magnetic system within the BG
distributions (\ref{BG} ) in the $M-U$ diagram
(magnetization-internal energy). White-region: stable states with
$\chi _{T}>0 $ and $M\cdot H>0$; Gray region: metastable states with
$\chi _{T}>0$ and $M\cdot H<0$; Light-gray region: unstable states
with $\chi _{T}<0$ and $M\cdot H<0$. The critical point \textbf{C}
with $\beta =\beta _{c}$ and $H=0$ is a state of \emph{marginal stability}
for BG distributions. Panel (b) Behavior of some
typical isotherms in the $H-M$ diagram.}
\label{scheme.eps}
\end{figure}

Let us enter into the central discussion of this work: the implication of the extended EFT (\ref{efdt}) on critical phenomena. Our interest is to analyze the thermodynamic behavior of a magnetic system when the same one is immersed into an environment with constant inverse temperature $\beta $; at the same time, it is put under the influence of an external magnetic field, $H$, that \textit{experiences a non-vanishing magnetic feedback effect} $\left\langle \delta H\delta M\right\rangle$. This kind of environmental influence naturally appears when the magnetic field $H^{s}$ associated with the system magnetization $M$ affects the source of the external magnetic field $H$ and induces the existence of non-vanishing correlations $\left\langle \delta H\delta M\right\rangle$. For the sake of simplicity, it has been assumed that the external magnetic field $H$ is weakly perturbed by the magnetization fluctuations, so that, this effect can be considered within the linear approximation $H=H_{0}-\lambda \delta M/N$; where $N$ is the system size, $H_{0}=\left\langle H\right\rangle $, the expectation value of the external magnetic field $H$, and $\lambda $, an effective coupling constant that characterizes the system-environment magnetic interaction. In the case of $\lambda =0$, it considers the equilibrium situation described by BG distributions (\ref{BG}). Conversely, a value $\lambda \neq 0$ only ensures the constancy of the external magnetic field $H$ in average sense. The application of the extended EFT (\ref{efdt}) to this type equilibrium situation leads to the following relation for the susceptibility $\chi_{T}$:
\begin{equation}
\beta \chi _{T}=\beta ^{2}\left\langle \delta M^{2}\right\rangle
-\beta ^{2}\chi _{T}\left\langle \delta H\delta M\right\rangle ,
\end{equation}%
and therefore:
\begin{equation}
\beta \left\langle \delta M^{2}\right\rangle =\frac{\chi
_{T}}{1+\lambda
\chi _{T}/N}\Rightarrow \left\langle \delta H^{2}\right\rangle =\frac{%
\lambda }{\beta N}\frac{\lambda \chi _{T}/N}{1+\lambda \chi _{T}/N}.
\label{flu1}
\end{equation}%
Admitting an extensive growth of the susceptibility $\chi _{T}$ with
$N$ as $\chi _{T}\propto N$, it is possible to verify that the
self-correlation of the external magnetic field behaves as
$\left\langle \delta H^{2}\right\rangle \propto 1/N$. From the
thermodynamic viewpoint, $\lambda \not=0$ does not differ from
$\lambda =0$ when $N\rightarrow \infty $. However, the positive
character of the self-correlation function $\left\langle \delta
M^{2}\right\rangle $ now leads to the following condition of
thermodynamic stability:
\begin{equation}
\lambda +N/\chi _{T}>0.  \label{stab2}
\end{equation}
According to Eq.(\ref{flu1}), the self-correlation function of
magnetization $\left\langle \delta M^{2}\right\rangle $
\textit{remains finite} when $\chi _{T}\rightarrow \infty $ as long
as the coupling constant $\lambda $ takes a positive value.
Consequently, there is no divergence of the correlation length $\xi
$ for this special equilibrium situation. Moreover, the
consideration of a coupling constant with $ \lambda >0$ enables the
access to \textit{diamagnetic states} $ \chi _{T}<0$ that exist
below the critical temperature, which are schematically represented
in Fig. \ref{scheme.eps}.

Let us suppose that the magnetic system is initially set at a state
\textbf{O} in the line with zero magnetization at
Fig. \ref{scheme.eps}.a, where $\beta <\beta _{c}$ and $H_{0}=0$.
Considering that the environmental inverse temperature $\beta$ is
increased with and without the magnetic feedback effect
$\left\langle \delta H\delta M\right\rangle$, a relevant question
emerges: \textit{What is the system behavior under these two
external influences}? The states with zero magnetization, $M=0$,
inside the region with $\chi _{T}<0$ are unstable within the BG
distribution (\ref {BG}). In this case, the self-correlation
function $\left\langle \delta M^{2}\right\rangle $ diverges for
$\lambda=0$ when the system approaches the critical point
$\textbf{C}$ (where $\chi _{T}\rightarrow \infty $). For $\beta
>\beta _{c}$, the system is forced to move along the stable
symmetric curves $\alpha $ or $ \alpha ^{\prime }$ with
non-vanishing magnetization, where it undergoes a spontaneous
symmetry breaking. Conversely, diamagnetic states $\chi _{T}<0$ turn
stable choosing a positive value of the coupling constant $\lambda
$. In this latter situation, the self-correlation function
$\left\langle \delta M^{2}\right\rangle $ does not diverge when the
system approaches the critical point $\textbf{C}$. For $\beta>\beta
_{c}$, the system remains on the line with zero magnetization $M=0$
inside the region with $\chi _{T}<0$, which implies a
\textit{suppression of the spontaneous symmetry breaking}. Thus, the
system critical behavior can be modified by the environmental
influence. Noteworthy that the incidence of non-vanishing feedback
effect $\left\langle \delta H\delta M\right\rangle$ can also force
the divergence of the self-correlation function $\left\langle \delta
M^{2}\right\rangle $ at an arbitrary state \textbf{P} inside the
region with negative magnetic susceptibility. The occurrence of this
phenomenon demands a coupling constant $\lambda \equiv -N/\chi
_{T}\left( \mathbf{P}\right)
>0$. Thus, the simultaneous divergence of the correlation length
$\xi $ and the response function $\chi _{T}$ is only admissible if
$\lambda=0$, that is, in the framework of BG distributions
(\ref{BG}).

The previous analysis can be verified through appropriate Monte
Carlo (MC) simulations. For this purpose, we consider the Ising model on
a square lattice $L\times L$ with periodic boundary conditions,
whose internal energy $U=-\sum_{\left\langle ij\right\rangle }\sigma
_{i}\sigma _{j}$ and magnetization $M=\sum_{i}\sigma _{i}$, where
$\sigma _{i}=\pm 1$ and $ \langle ij\rangle $ denotes the nearest
neighbors sites. As the case of conventional Metropolis importance
sample \cite{metro}\ based on the BG distributions (\ref{BG}), the
present situation with $\left\langle \delta H\delta M\right\rangle
\not=0$ can be implemented through the following acceptance
probability $p$ for a MC step from the state $(U,M)$ to $(U+\Delta U,M+\Delta
M)$:
\begin{equation}
p=\min \left\{ \exp \left[ -\beta \left( \Delta U+H\Delta M\right)
\right] ,1\right\} ,  \label{u.ap}
\end{equation}%
where the magnetic field $H$ has an implicit dependence on the
magnetization fluctuations.
\begin{figure}[t]
\begin{center}
\includegraphics[
width=4.0 in ]{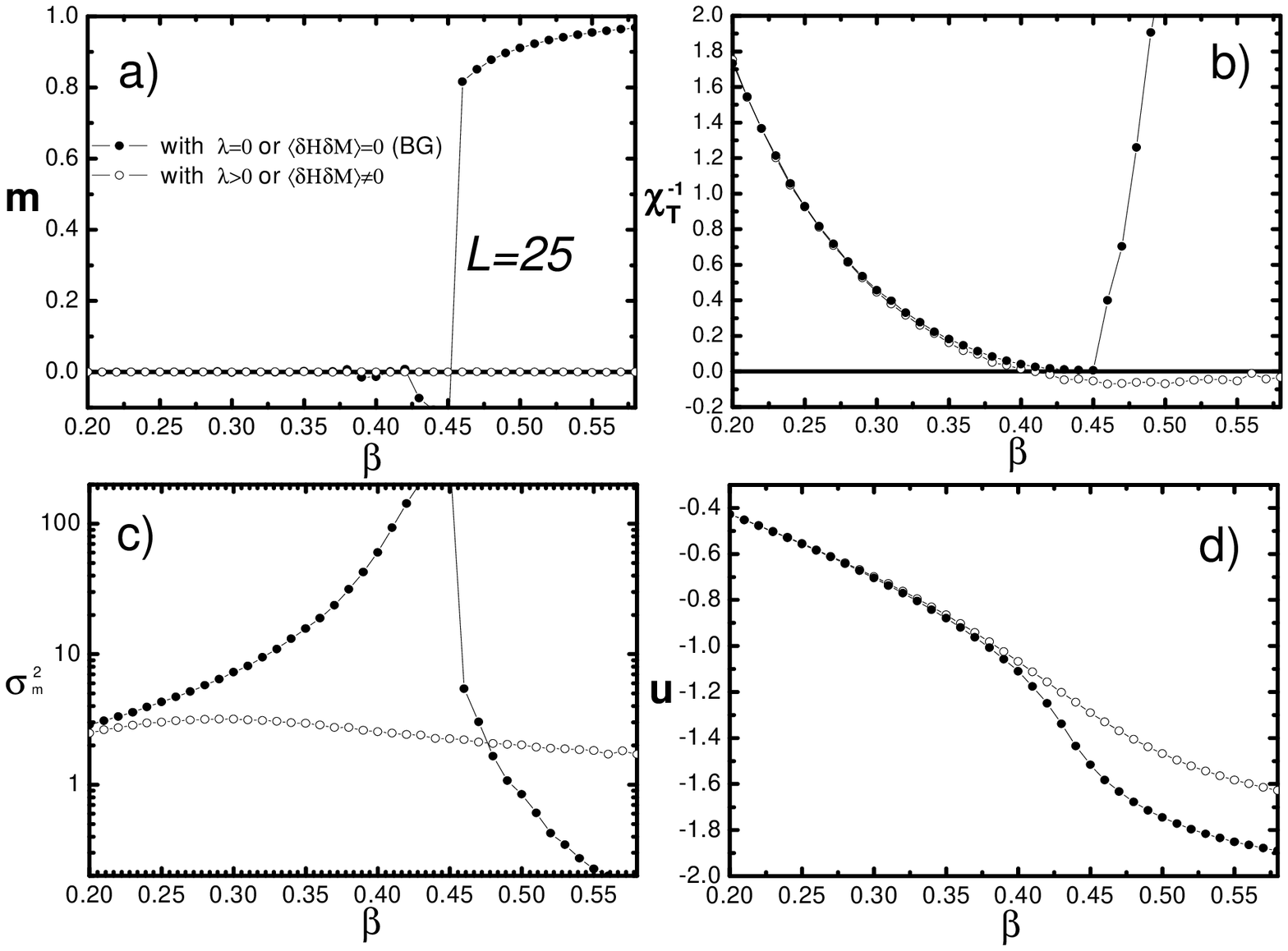}
\end{center}
\caption{Inverse temperature dependence of the thermodynamic
behavior of the 2D Ising model with $L=25$ and $H_{0}=0$ with (open
circles) and without (black circles) a magnetic feedback effect
$\left\langle \delta H\delta M\right\rangle$ obtained from the Monte
Carlo simulations with $n=10^{5}$ iterations per point.}
\label{study.eps}
\end{figure}
The magnetization dispersion $\sigma _{m}^{2}=\left\langle \delta
M^{2}\right\rangle /N$ can be employed to obtain the inverse
susceptibility per particle as $\bar{\chi}_{T}^{-1}=N\chi
_{T}^{-1}=\left( 1-\lambda \beta \sigma _{m}^{2}\right) /\beta
\sigma _{m}^{2}$. Along the Monte Carlo simulations with
$\left\langle \delta H\delta M\right\rangle\neq0$, the value of the
coupling constant $\lambda$ is redefined for each calculated point
using the optimal value $\lambda _{opt}=\sqrt{1+\left(
\bar{\chi}_{T}^{-1}\right) ^{2}}-\bar{\chi}_{T}^{-1}$, which
minimizes the total dispersion $\sigma ^{2}=\left\langle \delta
M^{2}\right\rangle /N+N\left\langle \delta H^{2}\right\rangle $ and
satisfies the stability condition (\ref{stab2}).

A comparative study of Monte Carlo simulations with and without a
magnetic feedback effect $\left\langle \delta H\delta
M\right\rangle$ is shown in FIG.\ref{study.eps}. According to
results that we show in FIG.\ref{study.eps}.c, the magnetization
dispersion $\sigma _{m}^{2}$ diverges at the critical point $\beta
_{c}\left( L=25\right) \simeq 0.41$ in Monte Carlo simulations with
coupling constant $\lambda=0$. Conversely, the same quantity remains
finite when $\lambda>0$, which indicates a suppression of the
\textit{long-range order} associated with the divergence of
correlation length $\xi \rightarrow \infty $. Simultaneously, there
exist a suppression of the spontaneous symmetric breaking, which is
manifested in FIG.\ref{study.eps}.a and FIG.\ref{study.eps}.b as a
persistence of diamagnetic states, $\chi_{T}<0$, with $\left\langle
M\right\rangle =0$ for $\beta>\beta _{c}$. Remarkably, the absolute
values $\left|\bar{\chi }_{T}\right|$ of these negative
susceptibilities are very large. The suppression of the spontaneous
symmetric breaking is also manifested as a \textit{bifurcation} of
caloric curves and the inverse isothermal susceptibility $\bar{\chi
}_{T}^{-1}$ at the critical point \textbf{C} shown in
FIG.\ref{study.eps}.b and FIG.\ref{study.eps}.d, a fact that gives
evidences to the different thermodynamic behavior of this system
inside the region with negative susceptibilities.

\section{Concluding remarks}
As already evidenced, the extended EFT (\ref{efdt}) reveals some new insights in the understanding on statistical physics. This theorem clearly distinguishes two different type of thermodynamic properties for a real system: the \emph{intrinsic properties} and the \emph{relative ones}. Examples of intrinsic properties are \emph{the response functions} (\ref{response}), \emph{the state equations} (\ref{var.deriv}), as well as any other information derived from the thermodynamics of the isolated system; in particular, from the knowledge of the entropy $S$. Conversely, the relative thermodynamic properties depend on the environmental influence acting in a concrete equilibrium situation, as example, any information concerning to the system stability and its fluctuating behavior.
The divergence of correlation length $\xi$ and its associated critical behaviors (spontaneous symmetry breaking, critical opalescence, etc.) take place at states with a \emph{marginal stability} (see in Fig.\ref{scheme.eps}.a). Since stability conditions, e.g.: Eq.(\ref{stab2}), crucially depend on the environmental influence, \emph{any critical behavior is always a relative phenomenon}. Hence, the divergence of any response function component $\mathcal{R}^{ij}\rightarrow\infty$ is \emph{not an authentic critical behavior}, but an \emph{intrinsic phenomenon} that has nothing to do with the divergence of correlation length $\xi$. 
The above remarks evidence that the consideration of BG distributions (\ref{BG}) can potentially lead to some incorrect predictions about the system thermodynamical behavior; overall, when one assumes these equilibrium distribution functions outside the context of its traditional applications. This could be the case of collective phenomena involving systems with long-range interactions as non-screened plasmas and astrophysical systems \cite{Dauxois}, or, the small and mesoscopic systems as the case of nuclear multifragmentation \cite{Moretto}, where the existence of states with anomalous response is almost a rule rather than an exception.

Before end this section, we would like comment three possible application frameworks of our approach, which could deserve a special attention in future works. Previously, this framework of fluctuation theory have been successfully applied to overcome slow sampling problems in Monte Carlo simulations associated with the incidence of discontinuous phase transitions \cite{vel.mc,vel.mc2}. Analogously, the suppression of divergence of correlation length $\xi$ at the critical point could be also employed to design new Monte Carlo algorithms to deal with difficulties associated with continuous phase transitions. The second application framework are the \emph{quantum theories at finite temperature} \cite{Kapusta}, which exploit the analogy between the evolution operator $\hat{T}=e^{-it\hat{H}/\hbar}$ and the statistical operator $\hat{\omega}=e^{-\beta\hat{H}/k_{B}}$. In high energy physics, these formulations have been applied to study processes that took place at the early universe as a whole, as example, spontaneous symmetry breaking of electro-weak interactions. Since these processes are special cases of critical phenomena, the application of present framework of fluctuation theory could reveal new insights in their understanding. Finally, the analogy between electric and magnetic systems strongly suggests the existence of anomalous \emph{dia-electric states} (electric counterpart of diamagnetic states) below critical temperature of a ferro-electric system, which should not be detected using experimental techniques based on Boltzmann-Gibbs distributions (\ref{BG}). We think that the ideas discussed in this work could inspire new experimental techniques to clarify the existence (or nonexistence) of these anomalous states. As discussed elsewhere \cite{Ginzburg}, the presence of a hypothetic dia-electric medium favors the formation of Cooper-pairs, a mechanism that triggers the development of \emph{high-temperature superconductivity}.

\section*{Acknowledgments}
L Velazquez thanks the financial support of
CONICYT/Programa Bicentenario de Ciencia y Tecnolog\'{i}a PSD
\textbf{65} (Chilean agency).


\section*{References}
\begin{thebibliography}{0}
\bibitem{Stanley} Stanley H. E., \emph{Introduction to phase transitions and
critical phenomena} (Clarendon Press, Oxford, 1971).

\bibitem{Gold} Goldenfeld N., \emph{Lectures on phase transitions and
critical phenomena, Frontiers in physics 85} (Perseus Books Publishing, L.L.C., 1992).

\bibitem{vel.unc} Velazquez L. and Curilef S., J. Phys. A: Math. Theor. \textbf{42} (2009) 095006.

\bibitem{vel.unc11} Velazquez L. and Curilef S., J. Phys. A: Math. Theor. \textbf{42} (2009) 335003.

\bibitem{vel.unc12} Velazquez L. and Curilef S., J. Stat. Mech.: Theo. Exp. (2009) \textbf{P03027}.

\bibitem{vel.unc2} Velazquez L. and Curilef S., J. Stat. Mech.: Theo. Exp. (2010) \textbf{P12031}.

\bibitem{vel.unc3} Velazquez L. and Curilef S., Mod. Phys. Lett. B \textbf{23} (2009) 3551.

\bibitem{Reichl} Reichl L.E., \textit{A modern course in Statistical
Mechanics} (Univ. Texas Press, Austin, 1980).

\bibitem{Gross} Gross D. H. E., \textit{Microcanonical thermodynamics: Phase
transitions in Small systems, 66 Lectures Notes in Physics} (World scientific, Singapore, 2001).

\bibitem{thir} Thirring W., Z. Phys. \textbf{235} (1970) 339;
  \textit{Quantum Mechanics of large systems} (Springer, 1980) Chapter 2.3.

\bibitem{Dauxois} \textit{Dynamics and Thermodynamics of Systems with Long Range
Interactions} Eds. Dauxois T., Ruffo S., Arimondo E. and Wilkens M., (Springer, New York, 2002).

\bibitem{Moretto} Moretto L. G., Ghetti R., Phair L., Tso K. and Wozniak G. J., Phys. Rep. \textbf{287} (1997) 250.

\bibitem{metro} Metropolis N., Rosenbluth A. W., Rosenbluth M. N.,
Teller A. H. and Teller E., J. Chem. Phys. \textbf{21} (1953) 1087.


\bibitem{vel.mc} Velazquez L. and Curilef S., J. Stat. Mech.: Theo. Exp. (2010) \textbf{P02002}.

\bibitem{vel.mc2} Velazquez L. and Curilef S., J. Stat. Mech.: Theo. Exp. (2010) \textbf{P04026}.

\bibitem{Kapusta} Kapusta J. I., \textit{Finite-temperature field theory} (Cambridge University Press, 1989).

\bibitem{Ginzburg} Ginzburg V. L., Rev. Mod. Phys. \textbf{76} (2004) 981.

\end{thebibliography}
\end{document}